\documentclass[twocolumn]{aastex701}
\usepackage{bm}
\def\lsim{~\rlap{$<$}{\lower 1.0ex\hbox{$\sim$}}}

\def\gsim{~\rlap{$>$}{\lower 1.0ex\hbox{$\sim$}}}

\def\V{\mathcal{V}}
\def\u{{\bf U}} 

\def\N{\mathcal{N}}

\begin{document}

\title[Gravitational-Electric Polarization]{Gravitational-Electric Polarization as a Probe of Dark Matter and Modified Gravity}

\author[orcid=0000-0001-9829-7727,gname=Nirupam,sname=Roy]{Nirupam Roy} 
\affiliation{Department of Physics, Indian Institute of Science, Bangalore 560012, India}
\affiliation{Department of Physics, New Mexico Institute of Mining and Technology, Socorro, NM 87801, USA}
\email[show]{nroy@iisc.ac.in}

\begin{abstract}

Self-gravitating astrophysical plasmas naturally achieve a state of global electrical polarization, known as the Bally-Harrison effect, where an induced electric field counteracts the preferential thermal escape of electrons. In this work, we revisit the phenomenon of gravitational-electric polarization in astrophysical plasmas. By accounting for the dominant role of dark matter and comparing results across modified gravity frameworks, including MOND and MOG, we provide new constraints on the global charge-to-mass ratios of galaxies and clusters. We demonstrate that the effective charge-to-baryonic-mass ratio $Q/M_{\text{bar}}$ is enhanced by a factor of 10--30 at the virial radii relative to purely baryonic predictions. By coupling gravitational polarization to galactic rotation, we derive a structurally linked seed field that reaches $\sim 10^{-23}$~G in high-redshift proto-galaxies, sufficient for rapid dynamo saturation. We demonstrate that the distinct spatial signatures of these fields across different gravity theories provide a potential observational probe of the dark sector in the early universe. This enhancement may have significant implications in inferring properties of the intracluster medium and in determining the primordial seed magnetic field. The  distinct radial and mass-dependent scaling laws predicted for each paradigm also provide a plausible diagnostic to distinguish between the presence of invisible mass and modifications to the gravitational law. 

\end{abstract}



\section{Introduction}
\label{sec:intro}

The assumption of local and global electrical neutrality of the baryonic matter is a fundamental tenet in most cosmological and astrophysical models. However, in any self-gravitating system composed of plasma, the difference masses of the ions and the electrons lead to a non-trivial equilibrium state. As first noted by \citet{Bally1978}, electrons in a gravitational potential possess thermal velocities higher than those of protons, allowing them to preferentially escape or occupy higher-energy states. This ``leakage'' continues until a net positive charge accumulates, creating an inward-directed electric field that counteracts the outward thermal pressure. The resulting, albeit weak, polarization ensures that the system reaches a steady state where the electrical and gravitational potentials are coupled, transforming stars, galaxies, and clusters into globally polarized macro-structures.

In the decades since this mechanism was proposed, our understanding of the gravitational landscape has changed significantly with clear observational confirmation of missing mass on galactic and extragalactic scales. While the original Bally-Harrison model focused on a purely baryonic framework, the presence of Cold Dark Matter (CDM) halos provides a much deeper gravitational ``trap''. Because dark matter interacts only gravitationally, it contributes to the potential well, while only the baryonic component can respond to and generate the resulting electromagnetic fields. This disconnect implies that the degree of polarization in a $\Lambda$CDM universe is significantly higher than initially estimated, with profound implications for the dynamics of the intracluster medium (ICM) and the propagation of charged cosmic rays.

Recent developments in relativistic astrophysics have revisited these charge asymmetries, particularly in the context of compact objects and galactic nuclei. \citet{Zajacek2019} demonstrated that macroscopic bodies such as black holes can maintain a net charge-to-mass ratio of approximately $150$ C/$M_{\odot}$. While this charge is negligible in terms of the spacetime metric, it is dynamically significant for the behavior of ionized gas (e.g. plasma accretion and jet formation) in the vicinity of supermassive black holes. Furthermore, \citet{Padilla2023} suggest that such charge imbalance and gravitational-electric polarization in hydrostatic equilibrium could provide the necessary ``seed'' fields ($B \sim 10^{-30}$ G) in the early universe, which are later amplified by dynamo mechanisms. Despite these advances, a systematic comparison between the polarization signatures in standard gravity versus modified gravity frameworks remains a critical gap in the literature. 

The emergence of Modified Newtonian Dynamics (MOND) \citep{Milgrom1983} and Scalar-Tensor-Vector Gravity (MOG) \citep{Moffat2006} as alternatives to the standard cold dark matter paradigm necessitates a re-evaluation of the Bally-Harrison effect. In these models, the ``missing mass'' is effectively replaced by a modification of the law of inertia, the gravitational law, or a boosted effective gravitational constant. For a MONDian system, the polarization is expected to exhibit a unique radial dependence in the low-acceleration limit, whereas MOG predicts an extreme amplification of the charge-to-mass ratio to compensate for the absence of a dark halo. These distinctly different predictions offer a unique observational window: the electrostatic support of hot gas in galaxy clusters may serve as a means to distinguish between invisible mass and modified gravity laws.

In this paper, we provide a rigorous update to the Bally-Harrison calculation by explicitly incorporating the Dark Matter fraction. We then derive the expected polarization signatures for MOND and MOG, identifying distinct scaling laws for each. We then consider the resultant seed magnetic field and its implications for galactic scale dynamo. We conclude by discussing the observational consequences, including the puffing of galactic gas disks and potential discrepancies between X-ray and weak-lensing mass estimates in rich clusters.

\section{Cold dark matter paradigm} 
In the standard cosmological framework, the gravitational potential $\Phi$ of a virialized system is dominated by the CDM halo. While dark matter is collisionless and electrically neutral, its presence fundamentally alters the electrostatic equilibrium of the baryonic plasma. 

Considering a plasma of protons and electrons in hydrostatic equilibrium within a total gravitational potential $\Phi_{\text{tot}}(r)$, where $M_{\text{tot}}(r) = M_{\text{bar}}(r) + M_{\text{DM}}(r)$, the electron fluid must satisfy the momentum balance equation
\begin{equation}
n_e e \nabla \phi - \nabla P_e - n_e m_e \nabla \Phi_{\text{tot}} = 0
\end{equation}
where $\phi$ is the induced electric potential and $P_e$ is the electron pressure. Assuming an isothermal state $P_e = n_e k_B T$, and noting that the gravitational force on electrons is dominated by the requirement to maintain quasi-neutrality with heavier ions, the electric field $\mathbf{E} = -\nabla \phi$ required to prevent electron escape is given by:
\begin{equation}
\mathbf{E} = -\frac{m_e}{e} \nabla \Phi_{\text{tot}} \approx \frac{G M_{\text{tot}}(r) m_e}{e r^2}
\end{equation}
In the Bally-Harrison limit \citep{Bally1978}, based on earlier work on stellar atmospheric polarization \citep{Pannekoek1922, Rosseland1924}, the net charge $Q$ required to balance the total gravitational mass $M_{\text{tot}}$ is:
\begin{equation}
Q = \frac{G M_{\text{tot}} m_p}{k e}
\label{eq:total_charge}
\end{equation}
where $k = (4\pi\epsilon_0)^{-1}$ is the Coulomb constant. Note that, as argued also in the stellar context \citep{Rosseland1924, Eddington1926}, Debye shielding cannot neutralize this global charge imbalance if the scale length of the system $L \gg \lambda_D$, where  
\begin{equation}
\lambda_D = \sqrt{\frac{\epsilon_0 k_B T}{n_e e^2}}
\end{equation}
is the Debye length \citep{Bally1978}. Considering typical astrophysical parameters, this condition is satisfied for stars, galaxies, as well as galaxy clusters, with $Q/M_{\text{bar}} \sim 150$~C/$M_{\odot}$ for all self-gravitating objects.

In the CDM paradigm, since $M_{\text{DM}}$ contributes to the potential but not to the number of charge carriers, the charge-to-mass ratio relative to the visible baryonic mass $M_{\text{bar}}$ becomes:
\begin{equation}
\left( \frac{Q}{M_{\text{bar}}} \right)_{\text{CDM}} = \frac{G m_p}{k e} \left( 1 + \frac{M_{\text{DM}}}{M_{\text{bar}}} \right)
\end{equation}
Given the cosmological ratio $\Omega_{\text{DM}}/\Omega_b \approx 5.37$ \citep{Planck2020}, the induced charge is, on average, roughly six times higher than predicted by purely baryonic models. Based on recent estimates of the Milky Way mass \citep{Posti2019, Cautun2020, Karukes2020}, this enhancement factor is $\sim 6$ considering the extent of the baryon distribution, but can be as high as $\sim 20$ at the virial radius. For galaxy clusters, based on observations \citep{Giodini2009, Matz2014} and cosmological simulations \citep{Schaye2023, Nelson2024, Rohr2025}, the total-to-baryonic mass ratio is found to be significantly higher in low-mass groups or feedback-impacted cluster cores ($\sim 20 - 30$), decreasing to $\sim 6$ asymptotically at the virial radius as it approaches the cosmic mean. Thus, the presence of dark matter halos necessiates an order of magnitude increased estimation of the Bally-Harrison limit.

\section{Alternative models} While the $\Lambda$CDM paradigm explains the polarization enhancement via an invisible mass component, alternative gravity frameworks such as Modified Newtonian Dynamics (MOND) and Modified Gravity (MOG) seek to account for the observed dynamics by modifying the laws of gravitation. In these models, the gravitational potential $\Phi$ is effectively amplified in low-acceleration regimes without the necessity of a dark matter halo, which fundamentally alters the predicted Bally-Harrison equilibrium by decoupling the field strength from the presence of non-baryonic matter. The specific formulations of how these varying descriptions of the gravitation refine the electric polarization are evaluated in the following subsections.

\subsection{Modified Newtonian Dynamics} 
In the framework of Modified Newtonian Dynamics (MOND), the gravitational acceleration $a_g$ is modified below a characteristic scale $a_0 \approx 1.2 \times 10^{-10}$\,m\,s$^{-2}$, such that the true acceleration is related to the Newtonian value $a_N$ through the equation $a_g \mu(a_g/a_0) = a_N$. For the Bally-Harrison effect in the deep-MOND regime ($a_g \ll a_0$), which is applicable for the outer regions of galaxies and galaxy clusters, the acceleration is given by $a_g(r) = \sqrt{G M_{\text{bar}} a_0}/r$  \citep{Milgrom1983, Famaey2012}. Consequently, the induced electric field $E$ scales with the square root of the baryonic mass $M_{\text{bar}}$, rather than linearly with a total mass $M_{\text{tot}}$ as in the $\Lambda$CDM models. If the net induced charge required to maintain electrostatic equilibrium against the MONDian gravitational ``pump'' is $Q$, then
\begin{equation}
E(r) = \frac{Q}{4\pi\epsilon_0 r^2} = \frac{m_p}{e} \frac{\sqrt{G M_{\text{bar}} a_0}}{r}.
\end{equation}
On a characteristic system scale $R$, the specific charge-to-mass ratio is expressed as:
\begin{equation}
\frac{Q}{M_{\text{bar}}} = \frac{4\pi\epsilon_0 m_p}{e} R \sqrt{\frac{G a_0}{M_{\text{bar}}}}.
\end{equation}
This result indicates that in MOND, $Q/M_{\text{bar}}$ is not a constant ratio but has a $R/\sqrt{M_{\text{bar}}}$ scaling. This creates a distinct scenario different from the linear scaling predicted by dark matter halo models, providing a potential empirical test for the origin of galactic-scale electric potentials.

\subsection{Modified Gravity}
Modified Gravity (MOG), or Scalar-Tensor-Vector Gravity (STVG), replaces the dark matter hypothesis by introducing a repulsive vector field $\phi_\mu$ and treating $G$ and $\mu$ (the vector field mass) as dynamical scalar fields. The effective gravitational potential for a static, spherically symmetric source is 
\begin{equation}
\Phi_{\text{MOG}}(r) = -\frac{G_N M_{\text{bar}}}{r} \left[ 1 + \alpha - \alpha e^{-\mu r} \right].
\end{equation}
For $r \gg 1/\mu$, the gravitational ``pump'' is enhanced by the factor $(1+\alpha)$. To satisfy the equilibrium condition $e\mathbf{E} = -m_p \nabla \Phi$, the induced global charge $Q$ must scale with the enhanced potential. In MOG, the parameters are mass-dependent: $\alpha = \alpha_\infty \frac{M}{ (\sqrt{M} + \mathcal{E})^2 }$, where $M$ is the baryonic mass, $\alpha_\infty = 10$, and $\mathcal{E}^2 = 6.25\times10^8$~M$_{\odot}$ is the mass scale \citep{Moffat2006, Moffat2013}. This implies that the charge-to-mass ratio $Q/M_{\text{bar}}$ in MOG is a function of the total baryonic mass of the system, reaching a value of $\approx 1500$~C/$M_{\odot}$ for galaxies like the Milky Way and massive clusters where $\alpha \approx 10$.

\begin{figure*}
\begin{center}
\includegraphics[scale=0.35]{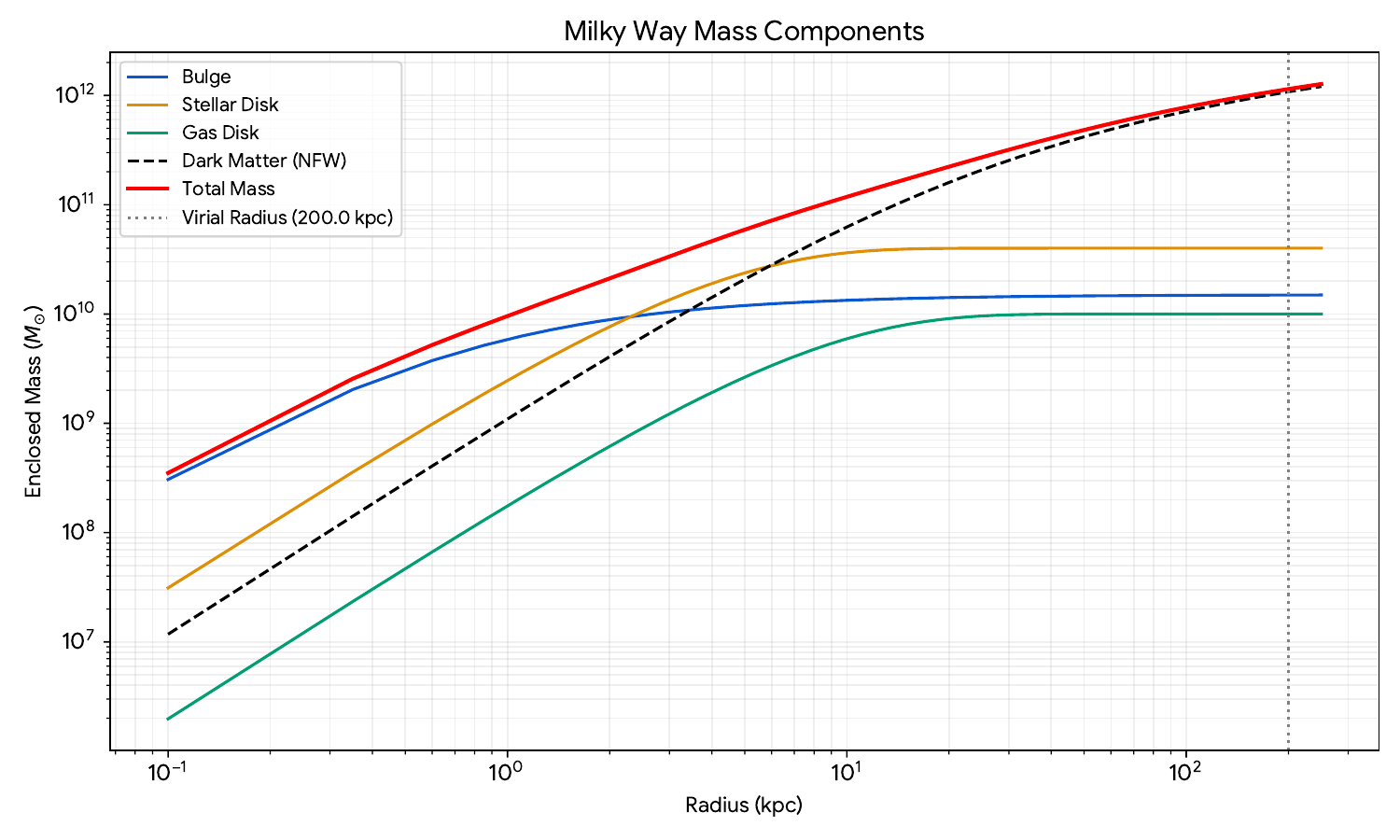}\includegraphics[scale=0.35]{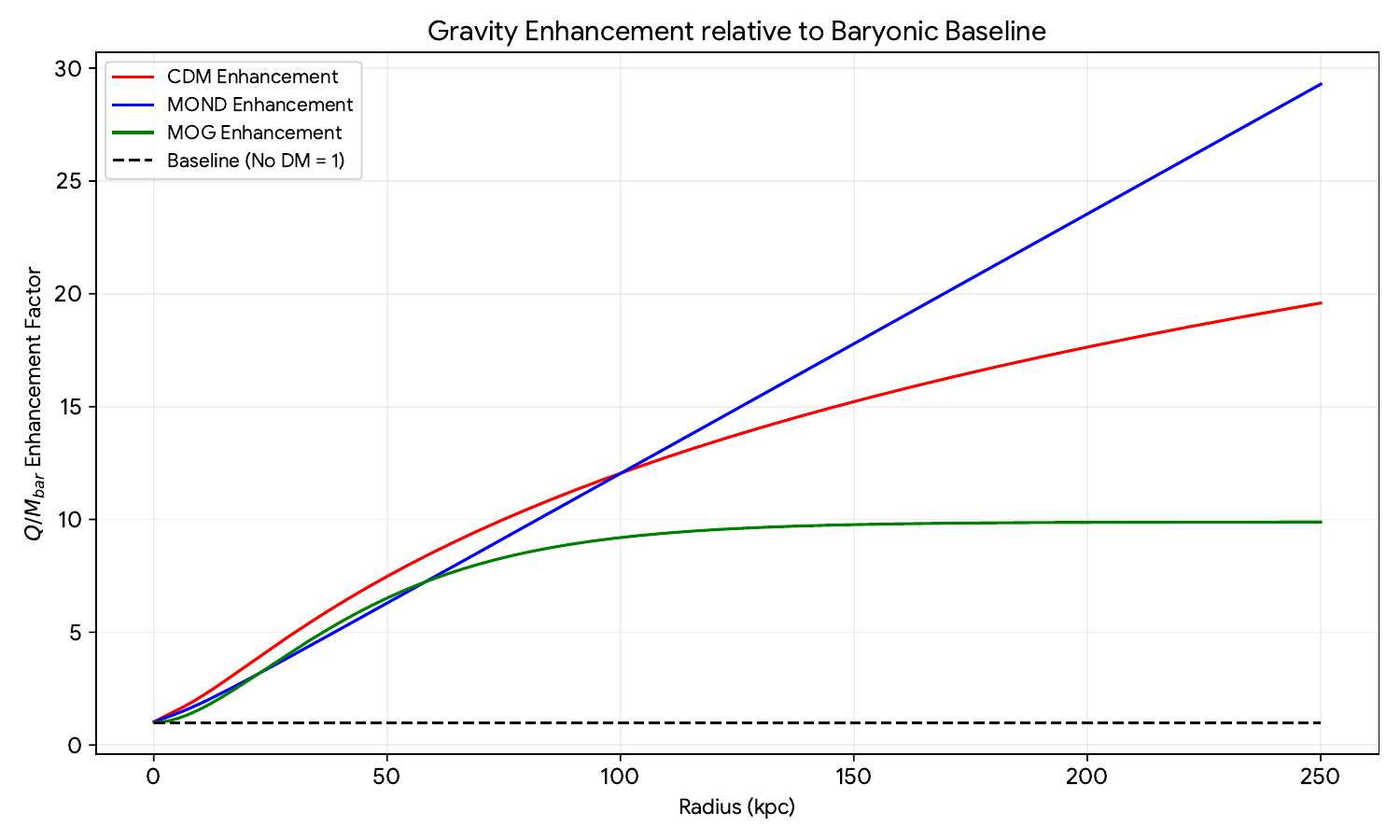}
\caption{{\it Left}: Radial distribution of mass components of a Milky Way like galaxy. {\it Right}: Gravitational enhancement factor as a function of galactocentric radius. Both MOND (blue) and MOG (green) show enhancement similar to that of the $\Lambda$CDM model in the inner part, but significant deviation in the outer part.}
\label{fig:pol_fig_1_2}
\end{center}
\end{figure*}

\begin{figure*}
\begin{center}
\includegraphics[scale=0.35]{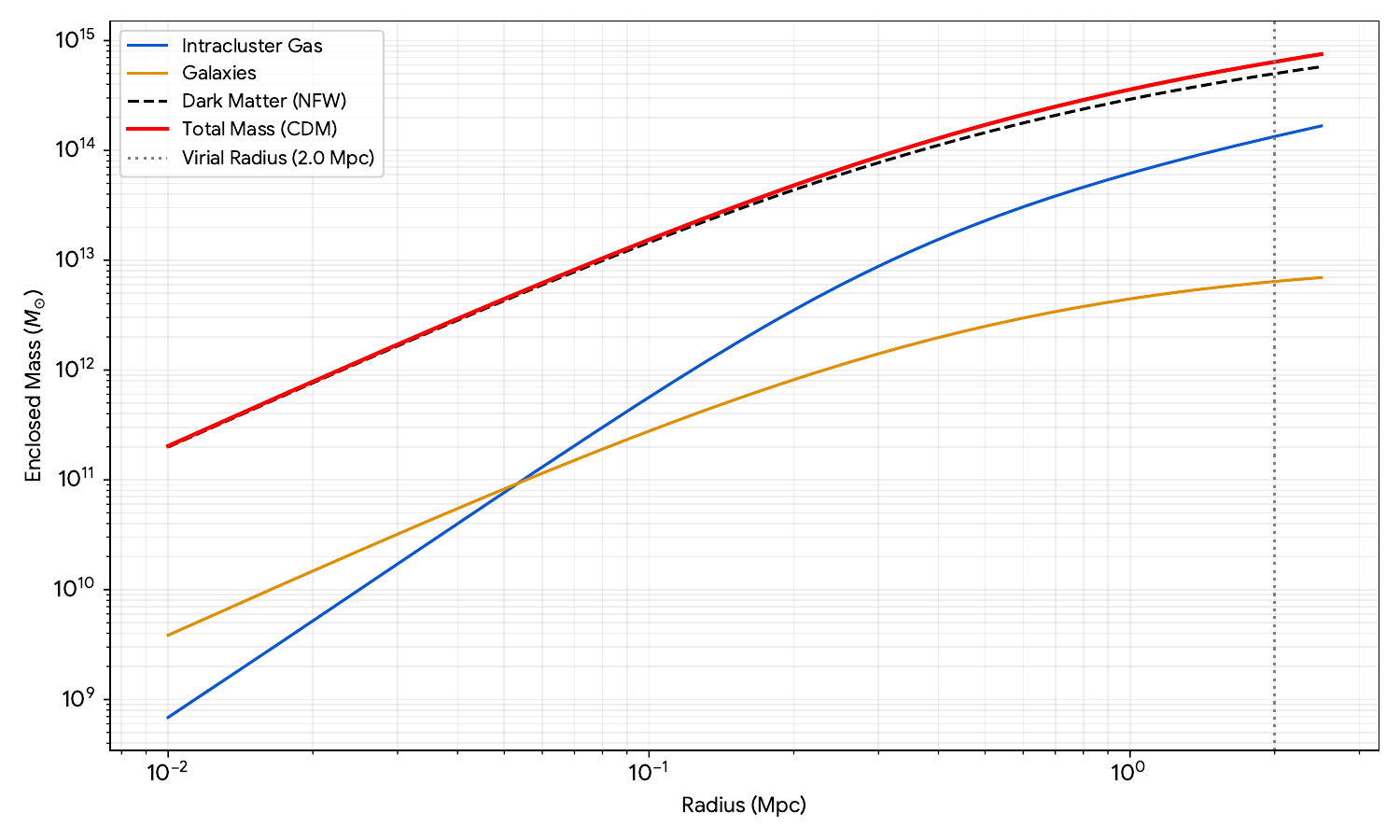}\includegraphics[scale=0.35]{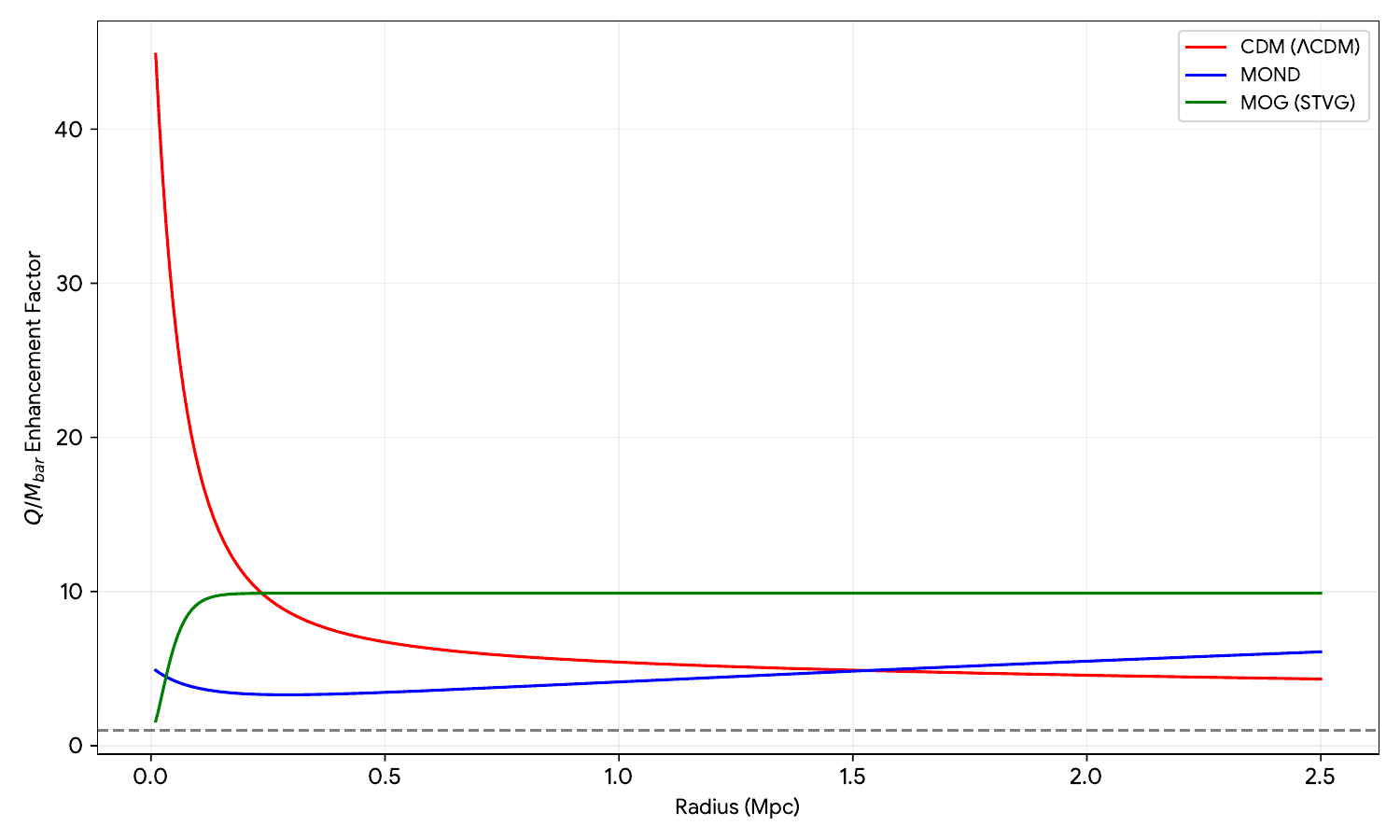}
\caption{{\it Left}: Radial distribution of model mass components of a galaxy cluster. {\it Right}: Gravitational enhancement factor as a function of radius.}
\label{fig:pol_fig_3_4}
\end{center}
\end{figure*}

For comparing the induced charge for different models, we consider the mass model of a Milky Way like galaxy. The stellar mass, consists of a compact central bulge modeled with a Hernquist profile and an exponential stellar disk, is taken to be $M_{\star} \approx 5.5 \times 10^{10} M_{\odot}$. The interstellar medium is represented by a gaseous disk component ($M_{\text{gas}} \approx 1 \times 10^{10} M_{\odot}$), resulting in a total baryonic budget of $M_{\text{bar}} \approx 6.6 \times 10^{10} M_{\odot}$. 

We evaluate the dynamics of this system under four distinct gravitational paradigms. In the ``No-DM'' baseline, the acceleration $g_{\text{bar}}$ is derived strictly from the Poisson equation using the observed baryonic distribution. Under the $\Lambda$CDM framework, these baryons are embedded within a spherical Navarro-Frenk-White (NFW) dark matter halo ($M_{\text{DM}} \approx 1 \times 10^{12} M_{\odot}$), which provides the missing dynamical mass. Alternatively, the MOND framework is implemented using the simple interpolation function $\mu(x) = x/(1+x)$, where $x = g/a_0$. Finally, the MOG model enhances the gravitational constant through a mass-dependent parameter $\alpha$ and a range parameter $\mu$, both scaled by the total baryonic mass $M_{\text{bar}}$.

As illustrated in Fig.~\ref{fig:pol_fig_1_2}, in the inner galaxy ($r < 5$ kpc), the enhancement factor $\eta \approx 1$ for all frameworks as baryons dominate the potential. In the galactic outskirts ($r > 15$ kpc), $\eta$ rises to $\approx 5 - 10$, where both MOND and MOG, even without assuming the presence of an NFW dark matter halo, predict values similar to that of the $\Lambda$CDM model. As expected, for the CDM model, the baryonic matter is concentrated in the central part of the halo, but the enclosed DM mass $M_{\text{DM}}$ has a slow growth, making the enhancement factor gradually increasing to $\approx 20$ at the virial radius ($R_{\text{vir}} \approx 200$~kpc). The MOG enhancement factor reaches an asymptotic value of $(1+\alpha) \approx 10$. On the other hand, deep MOND effect causes the enhancement of $Q/M_{\text{bar}}$ by a factor of $\approx 30$ at a similar radius in the outer part of the galaxy. Fig.~\ref{fig:pol_fig_3_4} shows the result of a similar analysis for the galaxy cluster. The left panel shows the radial profile of enclosed mass in galaxies, intracluster gas and DM halo for a Coma-like galaxy cluster model with a mass of  $M_{\text{gal}} \approx 1 \times 10^{13} M_{\odot}$, $M_{\text{ICM}} \approx 1.2 \times 10^{14} M_{\odot}$, and $M_{\text{DM}} \approx 5 \times 10^{14} M_{\odot}$ within a virial radius of $R_{\text{vir}} \approx 2.0$~Mpc. The right panel shows the enhancement of $Q/M_{\text{bar}}$ with respect to the Bally and Harrison baseline value of $\approx 1500$~C/$M_{\odot}$.

\section{Seed magnetic field}
\label{sec:sec3b}

The gravitationally induced charge density, $\rho_e$, coupled to the dynamics of the system, e.g. rotational velocity field, generates a polarization current $\mathbf{J} = \rho_e\mathbf{v}$. An ordered velocity field over a characteristic scale $L$ is, in turn, capable of generating a seed magnetic field $B$, estimated via Ampere’s law, is given by
\begin{equation}
B \approx \left( \frac{\mu_0 \epsilon_0 4\pi G m_p}{e} \right) \rho_{\text{tot}} v L\,.
\label{eqn:seedfield}
\end{equation}
Whereas computation of the exact value and spatial topology of the field should take into account the spatial variation of the induced charge density and the velocity field to derive $B$ from $\mathbf{J}$, the above expression can provide a back of envelope estimation of the order of magnitude of the seed field. 

For a typical baryon density of $10^{-21}$~kg~m$^{-3}$, $v \approx 200$~km~s$^{-1}$ and a characteristic scale of $L \sim 10$ kpc, the seed generated for the purely baryonic scenario will be $\sim 10^{-26}$~G. This is generally insufficient for $\mu$G saturation via standard dynamo amplification within the Hubble time. For $\Lambda$CDM, the DM halo increases the dynamical mass density $\rho_{\text{tot}}$ by a factor of $\sim 100$ and extends the field to the virial radius, thus making the lengthscale of interest $\sim 100$ kpc, and can generate a seed field of $\sim 10^{-23}$~G. Even considering typical parameters for a ``proto-galaxy'' with smaller $L$ (halo size $\sim 10$~kpc) as well as an order of magnitude lower DM halo mass at $z\approx3$, the seed magnetic field, following the scaling in equation~\ref{eqn:seedfield} will be $\gtrsim 10^{-22}$~G.

In the MOND/MOG frameworks, the effective gravitational force mimics the depth of a dark matter halo with mass as in the $\Lambda$CDM scenario, leading to a seed field of the same order. However, effective gravity in the deep MOND regime scales as $g \propto 1/r$. This leads to a more gradual radial decay of the induced field, producing a topology that is flatter and more persistent in the galactic outskirts than $\Lambda$CDM. MOG, on the other hand, introduces an enhancement factor of $(1+\alpha)$, but unlike the continuous DM density profile, MOG creates a characteristic ``plateau'' in the field strength due to the Yukawa range parameter $\mu$, producing a unique spatial signature. 

\subsection{Dynamo Implications and Observability}

We note that while a more conventional mechanism like Biermann battery, which generates seeds ($10^{-20}–10^{-23}$~G) via misaligned temperature and density gradients during shocks, is transient and dependent on localized thermodynamic fluctuations, this mechanism provides a steady-state, structurally linked seed that persists throughout the galaxy's evolution. The magnitude of the seed field is in the same range, but, unlike Biermann battery, expect to follow the DM or baryon distribution more closely.

Following the amplification frameworks of \citet{Arshakyan2009} and \citet{Schober2013}, these structured seeds are expected to provide the input necessary for the small-scale dynamo to reach microgauss ($\mu$G) saturation. We note that the seed field of $\sim 10^{-26}$~G for the baryon only scenario is of similar order as the magnetic fluctuations due to thermal noise and is susceptible to Ohmic diffusion on a shorter timescale before the turbulent eddies can fold them. For seed field strength $\gtrsim 10^{-22}$~G, the small-scale dynamo mechanism can be highly efficient, particularly in young galaxies, which can amplify weak seeds to microgauss levels on timescales significantly shorter than Hubble time \citep{Schober2013}. During the period of enhanced turbulence driven by intense star formation activities ($z\sim 2 - 3$), the typical turnover timescale is a few million years (taking the eddy size of $l\sim 100$~pc and turbulent velocity $v_{turb} \sim 50$~km~s$^{-1}$). This ensures that the estimated seed field can be amplified by the small-scale dynamo to $\mu$G saturation if the mechanism is operating for $\lesssim10\%$ of the time elapsed during this epoch. The evolutionary impact is determined by the spatial topology: CDM seeds track the extended dark matter halo, MOND seeds follow the baryonic distribution with a long-range enhancement, and MOG seeds exhibit a distinct plateau. 

Because the spatial scale of these seeds is inherently linked to the gravitational potential well \citep{Semikoz2009}, the magnetic field acts as a ``fossil record'' of the dark sector. Future high-resolution radio observations of high-redshift proto-galaxies could isolate these patterns, providing a novel, non-kinematic method to distinguish between dark matter halos and modified gravity theories. We leave a more detailed and quantitative analysis of these aspects for future work.

\section{Discussions}
\label{sec:sec4}

The analysis presented here reveals that while the baryonic component dominates the effective induced charge in high-density central regions of self-gravitating objects, a substantial divergence in the $Q/M_{\text{bar}}$ ratio emerges as one approaches the virial radius. The ratio $Q/M_{\text{bar}}$ is expected to have an order of magnitude enhancement, with a factor of $2 - 3$ variations between CDM, MOND and MOG predictions at $R_{\text{vir}}$ for both galaxies and clusters. 

The magnitude of this effective $Q/M_{\text{bar}}$ enhancement may have significant implications for the seeding of primordial magnetic fields. A factor of 10--20 increase in effective coupling will imply that fluctuations in the plasma during the Dark Ages may be far more efficient at generating magnetic seeds via effective polarization currents. The accelerated dynamo amplification of these seeds would lead to an early magnetization of the Intergalactic Medium (IGM). Such a magnetized environment may indirectly influence the redshifted 21~cm cosmological signal as well. An enhancement in the charge may also result in a significant momentum transfer cross-section, potentially leading to excess cooling anomalies observed in recent global 21~cm experiments.

The effect of the polarization on galactic scale is too weak and may be observationally inconsequential, although with an order of magnitude enhancement, it may lead to discernible difference in the flaring of the gas disk at the outer part. However, at the cluster scale, the $Q/M_{\text{bar}}$ profile may alter the inference of thermodynamic properties from the Intracluster Medium (ICM). As the observed X-ray temperature $T_X$ depends on the effective charge required to maintain hydrostatic equilibrium, any deviation in $Q/M_{\text{bar}}$ relative to the standard NFW density profile would lead to systematic biases in the inferred gas properties for MOND and MOG. 

It is important to note here that, while the net charge is a consequence of the (gradient of the) gravitational potential, which is deduced observationally from the rotation curve, the charge density is spatially modulated by the $M_{\text{DM}}/M_{\text{bar}}$ ratio. This creates a non-trivial profile that differs across $\Lambda$CDM, MOND and MOG. Thus, in addition to the overall order of magnitude enhancement, the coupling between the gravitational ``source'' and the electrical polarization varies fundamentally between the models. Specifically, the $Q/M_{\text{bar}}$ ratio differs by a factor of $2-3$ between models at the virial radius for both galaxies and galaxy clusters (and even more so for the central part of galaxy clusters). The influence of this on the associated electrostatic pressure and, in turn, on the thermodynamic profile of the ICM, may be important in interpreting the properties of X-ray-emitting gases out to a large radius. Similarly, the magnetic seeding being a dynamic evolutionary process, the spatially varying polarization profile may affect the magnetic seeding topology - a tangible difference that may be tested in future.

We note in passing that some millicharged dark matter (mDM) models may also alter the macroscopic electrical polarization. While globally neutral mDM will only cause an enhancement same as CDM, mDM with charge asymmetry will further alter the equilibrium electric field $\vec{E}$ and the charge-to-mass ratio $Q/M_{\text{bar}}$. For mDM with mass $m_\chi$ and fractional charge $\epsilon e$ to remain in equilibrium within the gravitational potential $\Phi$, the resulting charge-to-mass ratio is modified to $Q/M_{\text{bar}} \approx (G m_\chi / k\epsilon e)$ in the limit where mDM dominates the dynamics, i.e., if and when $(m_\chi / \epsilon m_p \gg 1)$. As $\epsilon \ll 1$, for ``WIMP-like'' mDM ($m_\chi \sim 100$~GeV), this predicts an amplification of the electric polarization by several orders of magnitude and may be used to rule out mDM parameter space.

\section{Summary and Conclusions}
\label{sec:sec5}

Traditionally, self-gravitating astrophysical plasmas achieve global electrical polarization through the Bally-Harrison effect, wherein an induced electric field counteracts the preferential thermal escape of electrons. The presence of dark matter or the application of modified gravity frameworks significantly deepens these potential wells, necessitating a rigorous re-evaluation of the resulting electrostatic equilibrium. By deriving the polarization signatures for $\Lambda$CDM, MOND, and MOG on galactic and cluster scales, we demonstrate that the effective charge-to-baryonic-mass ratio $Q/M_{\text{bar}}$ is enhanced by a factor of 10--30 at the virial radii, with each model exhibiting distinct radial and mass-dependent scaling laws. We show that this gravitational polarization generates steady-state seed magnetic fields of $\sim 10^{-23}$ G in early proto-galaxies, providing the necessary input for efficient small-scale dynamo amplification to $\mu$G levels. The resulting spatial topology acts as a fossil record of the underlying gravitational framework, offering a novel method to distinguish between $\Lambda$CDM, MOND, and MOG models through high-redshift observations. This study introduces a novel framework for distinguishing between invisible mass and modified gravity while highlighting the critical role of gravitational-electric polarization in seeding primordial magnetic fields,  influencing the 21~cm cosmological signal, and determining the properties of galaxy clusters. Future work extending this analysis to millicharged dark matter could provide a powerful diagnostic to further constrain the parameter space of dark sector interactions.

\begin{acknowledgments}
We sincerely thank the Scientific Editor Ethan Vishniac and the anonymous reviewer for their thoughtful comments and constructive suggestions, which helped improve the quality of this manuscript. NR acknowledges useful discussion with Rajaram Nityananda. NR also acknowledges support from the United States-India Educational Foundation through the Fulbright Program.
\end{acknowledgments}

\bibliography{references}{}
\bibliographystyle{aasjournalv7}



\end{document}